\documentclass[%
preprint,
 amsmath,amssymb,
 aps,
]{revtex4-2}

\usepackage{xcolor}
\usepackage{subfigure}
\usepackage{graphicx}
\usepackage{dcolumn}
\usepackage{bm}
\usepackage{hyperref}
\usepackage{epstopdf}

\begin{document}

\title{Modulating Spin Current Induced Effective Damping in $\beta-W/Py$ Heterostructures by a Systematic Variation in Resistivity of the Sputtered Deposited $\beta-W$ films}

\author{Soumik Aon}
 
\author{Sayani Pal, Subhadip Manna, Chiranjib Mitra}
 
 \author{Partha Mitra}

\affiliation{
 Department of Physical Sciences, Indian Institute of Science Education and Research, Kolkata, Mohanpur 741246, India 
}

\begin{abstract}
Utilizing the spin-induced pumping from a ferromagnet (FM) into a heavy metal (HM) under the ferromagnetic resonance (FMR) condition, we report an enhancement in effective damping in $\beta$- W/Py bilayers by systematically varying resistivity ($\rho_{W}$) of $\beta$-W films. Different resistivity ranging from 100 $\mu\Omega$-cm to 1400 $\mu\Omega$-cm with a thickness of 8 nm can be achieved by varying the argon pressure ($P_{Ar}$) during the growth by the method of sputtering. The coefficient of effective damping $\alpha_{eff}$ is observed to increase from 0.010 to 0.025 with $\rho_{W}$, which can be modulated by $P_{Ar}$. We observe a modest dependence of $\alpha_{eff}$ on the sputtering power ($p_{S}$) while keeping the $P_{Ar}$ constant. $\alpha_{eff}$ dependence on both $P_{Ar}$ and $p_{S}$ suggests that there exists a strong correlation between $\alpha_{eff}$ and $\rho_{W}$. It is thus possible to utilize $\rho_{W}$ as a tuning parameter to regulate the $\alpha_{eff}$, which can be advantageous for faster magnetization dynamics switching. The thickness dependence study of Py in the aforementioned bilayers manifests a higher spin mixing conductance ($g^{\uparrow\downarrow}_{eff}$) which suggests a strong spin pumping from Py into the $\beta$-W layer. The effective spin current ($J_{S(eff)}$) is also evaluated by considering the spin-back flow in this process. Intrinsic spin mixing conductance ($g^{\uparrow\downarrow}_{W}$) and spin diffusion length ($\lambda_{SD}$) of $\beta$-W are additionally investigated using thickness variations in $\beta$-W. Furthermore, the low-temperature study in $\beta$-W/Py reveals an intriguing temperature dependence in $\alpha_{eff}$ which is quite different from $\alpha_{b}$ of single Py layer and the enhancement in $\alpha_{eff}$ at low temperature can be attributed to the spin-induced pumping from Py layer into $\beta$-W.
\end{abstract} 
\maketitle
   \section{\label{sec:level1}Introduction}
Spin transport in metallic heterostructures is a growing topic of interest with the advent of new devices which utilize the phenomena like spin hall effect (SHE) \cite{hirsch,zhang,valenzuela2006direct,kimura2007room,brataas2012current,sinova2015}, spin transfer torque (STT) \cite{SLONCZEWSKI1996}, spin-orbit torque (SOT) \cite{miron2011perpendicular}, spin pumping \cite{Tserkovnyak2002FM, Brataas2002,lenz2004evidence,woltersdorf2005time,saitoh2006conversion, Brataas2008_scattering,ando2014dynamical}. The spin-orbit coupling \cite{vignale2010ten} is the most significant phenomenon causing these effects as it couples the charge $\&$ spin of an electron and controls the interconversion between charge and spin current. In the context of metallic multilayered devices, ferromagnetic resonance (FMR) induced spin pumping provides a reliable technique to inject pure spin current into the non-magnetic (NM) layer even without applying any charge current. Spin pumping is a phenomenon in which the magnetization precession in the ferromagnetic metal (FM) transfers the spin angular momentum into the NM layer, resulting in the generation of a spin current. According to the law of conservation of angular momentum, the spins ejected by spin current exert a torque on the FM layer \cite{Tserkovnyak2005_RevModPhys}, leading to the enhancement of the coefficient of effective damping ($\alpha_{eff}$), which can be evaluated by studying and analyzing the FMR absorption spectra.\\

Magnetization  dynamics of the FM/NM bilayer can be understood using the modified Landau-Lifsitz-Gilbert (LLG) equation as described by \cite{gilbert2004}.

\begin{equation}\label{eq:1}
\frac{d\vec{M}}{dt} = -\gamma(\vec{M}\times\vec{B}_{eff}) + \frac{\alpha_{b}}{M_{s}}(\vec{M}\times\frac{d\vec{M}}{dt}) + \frac{\gamma}{V}\vec{J}_{S}
\end{equation}
The conservative precessional term in a FM, where magnetization $\vec{M}$ precesses around the effective magnetic field $\vec{B}_{eff}$, is described by the first term on the right-hand side of Eq. \ref{eq:1}. However, the second term corresponds to the Gilbert damping with a damping coefficient $\alpha_{b}$ which quantifies the spin relaxation mechanism in bulk FM. The third term represents the spin pumping effect that results in a spin current injection from a FM layer into a NM layer. The spin current density $\vec{J}_{S}$ can be expressed as
 
\begin{equation} \label{eq:2}
\vec{J}_{S} = \frac{\hbar A}{4\pi M^{2}_{S}}g^{\uparrow\downarrow}_{eff} (\vec{M}\times\frac{d\vec{M}}{dt})
\end{equation}

which, leads to an additional damping ($\alpha_{SP}$), where the effective  damping is expressed as \cite{Tserkovnyak2002FM}: $\alpha_{eff} = \alpha_{b} + \alpha_{SP}$. Spin-orbit coupling (SOC) is a key factor in the enhancement of damping caused by spin pumping into the NM layer. The best candidates are the heavy metals (HM) and topological insulators (TI) \cite{Roja2016_TI} as they have high SOC. Platinum has been extensively explored as HM layer in FM/HM heterostructures for spin pumping experiments \cite{ando2009optimum,mosendz2010quantifying,ando2014dynamical,zhang2015role}. Among the other transition metals with higher SOC, tungsten (W) in its $\beta$-phase ($\beta$-W) shows the largest spin hall angle of $\approx$ 0.4 \cite{pai2012spin,hao2015beta,hao2015giant,demasius2016enhanced} (much larger than Pt \cite{mosendz2010quantifying,feng2012spin} and $\beta$-Ta \cite{liu2012spin}). $\beta$-W is characterized by high resistivity and has an A-15 crystalline structure. On the other hand, $\alpha$-W is characterized by its low resistivity and exhibits a small spin hall angle \cite{pai2012spin}. Recently, the dependence of spin pumping in different structural phases of tungsten in W/CoFeB bilayers has been studied at room temperature \cite{jhajhria2019}. Lu et al. \cite{lu2019enhancement} demonstrated an enhancement in spin mixing conductance by inserting a $\alpha$-W layer in between CoFeB/$\beta$-W heterostructure through interfacial phase engineering. There have not been many studies since then to optimize the W thin film in its $\beta$ phase with higher resistivity and investigate the spin pumping mechanism in HM/FM bilayers. This area of research holds promise for further exploration and understanding of spin transport properties in metallic heterostructures, potentially leading to the development of more efficient spintronics devices.\\

In our study, we investigate the spin pumping efficiency in $\beta$-W/Py bilayers by systematically varying the resistivity of $\beta$-W films by tuning the argon pressure ($P_{Ar}$) during the deposition through the dc sputtering technique, while maintaining the other parameters such as growth rate and sputtering power ($p_{s}$) constant. All deposited W films are single phase $\beta$-W with A15 crystal structure and resistivity increases with increasing $P_{Ar}$. We employ short-circuited CPW-based broadband VNA-FMR spectroscopy technique \cite{pal2022short,pal2023experimental} to measure the effective damping caused by spin pumping in $\beta$-W/Py bilayers. It has been observed that the resistivity of $\beta$-W ($\rho_{W}$) is strongly dependent on $P_{Ar}$ resulting in the enhancement in the effective damping $\alpha_{eff}$. Sputtering power $p_{s}$ dependence of $\alpha_{eff}$ is also discussed which is similar to fabrication kinematics of $P_{Ar}$. The other potential causes for the enhancement in $\alpha_{eff}$ are discussed in detail. The interface effect in $\beta$-W/Py is characterized by the effective spin mixing conductance $g^{\uparrow\downarrow}_{eff}$ which is extracted using the thickness variation of Py. The effective spin current $J_{S(eff)}$ is calculated while taking spin back-flow during spin pumping into consideration. Using the thickness dependence measurement of $\beta$-W, we are able to calculate the intrinsic/interfacial spin mixing conductance $g^{\uparrow\downarrow}_{W}$ and the spin diffusion length $\lambda_{SD}$. We investigate the temperature dependence of spin pumping in order to gain a deeper understanding of the damping mechanism in $\beta$-W. Our observation reveals an intriguing trend: as temperature decreases there is an increase in $\alpha_{eff}$ which can be explained by the torque-correlation model as previously reported \cite{kambersky1976ferromagnetic,Gilmore_2007_torque_correlation,Fahnle_2011_torque_correlation}. The device we employ in this paper is simple yet effective, offering valuable insights into the underlying physical mechanisms involved.

\subsection{\label{sec:level2}SAMPLE FABRICATION $\&$ EXPERIMENTAL DETAILS}
A series of W thin films (8 nm) were fabricated on $Si/SiO_{2}$ (300 nm) substrates at room temperature by dc magnetron sputtering technique equipped with both scroll and turbo pump, reaching a base pressure of $1\times10^{-6}$ Torr. The substrates of the dimension of 5 mm $\times$ 5 mm were placed in a sample holder and kept at a constant distance of 15 cm from the sputtering target. The dimension of the W target is 3 inches diameter $\times$ 3 mm thick (99.95\% pure). We used a straight deposition arrangement between the substrate and target to achieve uniform films. The W thin films were deposited at different argon pressure ($P_{Ar}$), varying systematically from 5.6 mTorr to 10 mTorr, with a constant sputtering power ($p_{S}$) = 150 watt. The deposition rate of W thin films at different $P_{Ar}$ decreased from 0.6 {\AA}/sec for $P_{Ar}$ = 5.6 mTorr to 0.3 {\AA}/sec for $P_{Ar}$ = 10 mTorr, recorded by a crystal monitor. The crystal structure of deposited W films was characterized using a Bruker x-ray diffractometer (XRD) with a Cu-$K_{\alpha}$ (1.54 {\AA}) source. The surface morphology and thickness profile of W films were studied using an atomic force microscope (AFM). The resistivity of W films was measured by the standard four-probe Van der Pauw technique. As no microfabrication was needed, Py films were then deposited on top of the entire $Si/SiO_{2}/W$ utilizing the thermal evaporation technique. The deposition rate of Py was fixed at 1 {\AA}/sec for all samples. Prior to each deposition, an ion milling has been performed using dry argon plasma to clean the surface of the substrates and to make the interface transparent. Py films were characterized by XRD and in-plane anisotropic magnetoresistance (AMR) measurement. \\  

\begin{figure}[h!]
\centering
 \includegraphics[width=0.4\textwidth]{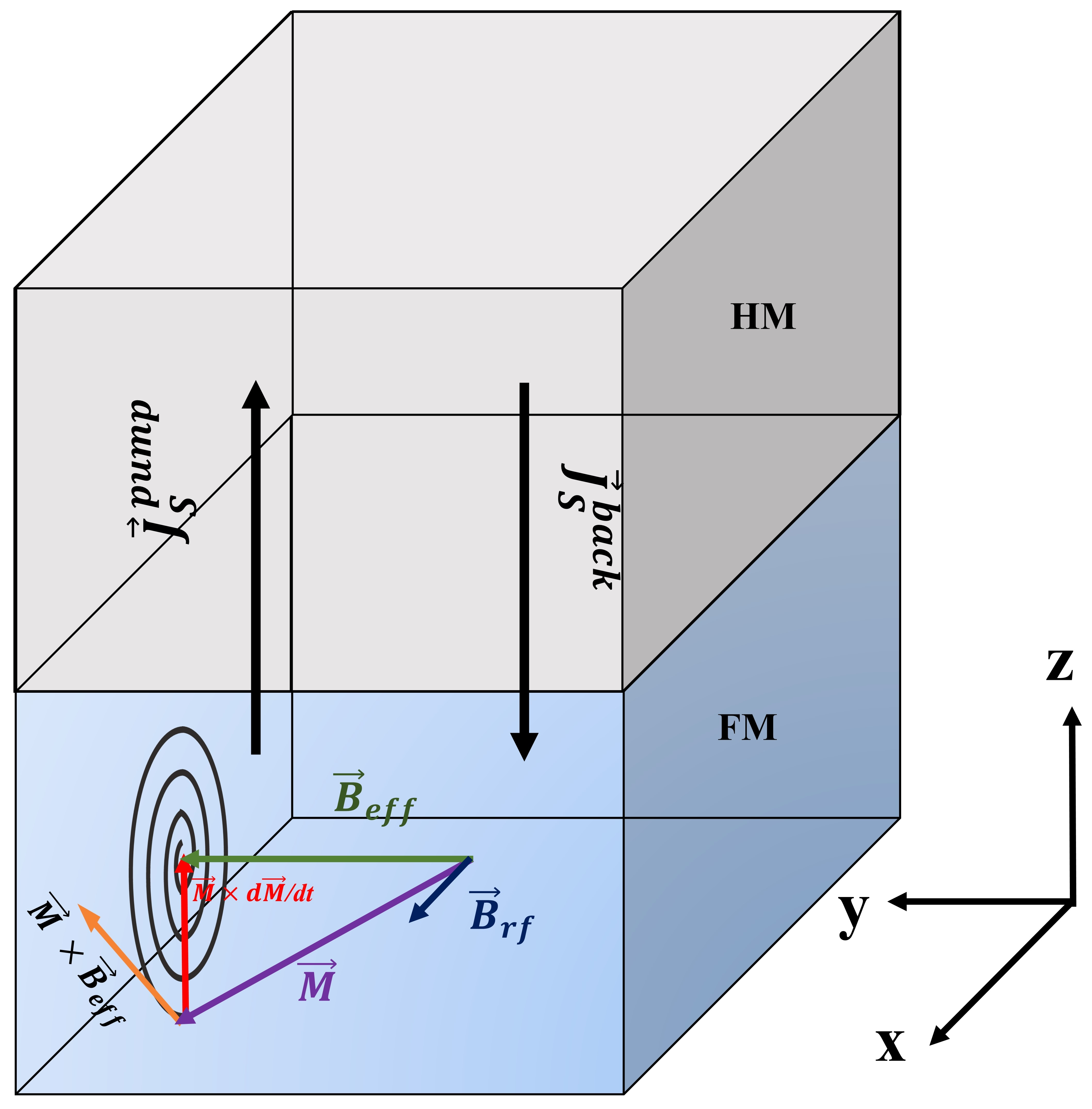}
\caption{Schematic of spin pumping in a HM/FM bilayer device. The device is placed on CPW upside down. Magnetization $\vec{M}$ in the FM layer precesses around the effective magnetic field $\vec{B}_{eff}$, generating a spin current that diffuses from the FM layer into the HM layer. For a realistic scenario, the pumped spin current  $J^{pump}_{S}$ generates a spin accumulation which results in a back-flow of the spin current denoted as $J^{back}_{S}$.}
\label{Figure1}
\end{figure}

For our experiment, we fabricated four different sets of samples: (i) $Si/SiO_{2}/\beta$-W (8nm, $P_{Ar}$)/Py (15 nm), where, $P_{Ar}$ was varied from 5.6 mTorr to 10 mTorr at constant sputtering power ($p_{s}$) of 150 watt. These samples, labeled as $\beta$-W ($P_{Ar}$)/Py (set A) were used to investigate the enhancement of effective damping at both room and low temperatures, (ii) dependence of effective damping on $p_{S}$ was studied using $\beta$-W(8nm, $p_{S}$)/Py(15 nm) (set B) bilayers with varying $p_{S}$ = 80, 150, 180, 250 and 330 watt, maintaining a constant $P_{Ar}$ = 6.7 mTorr, (iii) $\beta$-W($P_{Ar}$ = 6.7 mTorr)/Py ($t_{Py}$) (set C) bilayers with different Py thicknesses ($t_{Py}$ = 15, 20, 25, 30 nm) to evaluate $g^{\uparrow\downarrow}_{eff}$ and $J_{S(eff)}$ injected from FM layer into HM layer, (iv) $\beta$-W($P_{Ar}$ = 6.7 mTorr, $t_{W}$)/Py(15 nm) (set D) bilayers with different W thickness ($t_{W}$ = 2, 5, 8, 10, 15 nm) to estimate the  $g^{\uparrow\downarrow}_{W}$ and $\lambda_{sd}$ of $\beta$-W.\\
\begin{figure}[h!]
\centering
 \includegraphics[width=0.6\textwidth]{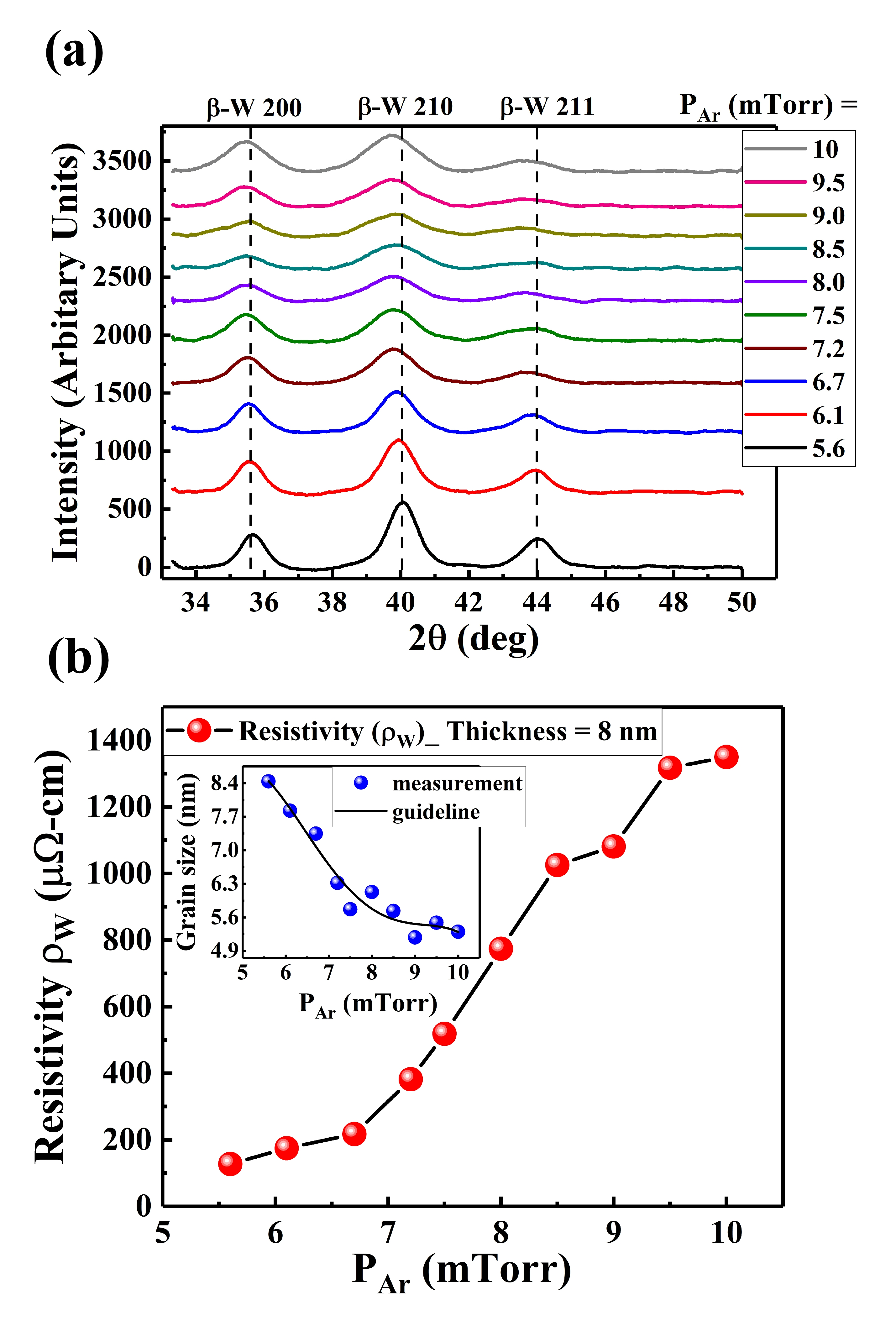}
\caption{(a) XRD of $\beta$-W thin films with various $P_{Ar}$. The peak positions with their relative peak intensities corresponding to $\beta$ phase of W are shown as vertical dotted lines, (b) Resistivity ($\rho_{W}$) as a function of $P_{Ar}$ (inset: Grain size of W films at different $P_{Ar}$).}
\label{Figure2}
\end{figure}

In our experimental setup, the samples were placed on top of the short-circuited coplanar waveguide structure (CPW) as described in \cite{pal2022short,pal2023experimental} and an in-plane external field (B) is applied, along with a radio frequency (rf) microwave magnetic field ($B_{rf}$) in a transverse direction. Due to the applied magnetic field, the magnetization starts to precess around the effective magnetic field ($B_{eff}$) at a frequency known as Larmor precession frequency which is shown in Fig. \ref{Figure1}. Absorption of electromagnetic energy occurs when the frequency of the $B_{rf}$ equals the Larmor frequency. There are two techniques that are used to get FMR spectra: sweeping frequency while maintaining a constant magnetic field or sweeping magnetic field while keeping a constant frequency. To obtain the ferromagnetic resonance (FMR) spectra, we employed the 2nd technique in our measurement where the magnetic field was varied. During the measurements, we employed frequencies between 3.5 GHz to 5.5 GHz, and the magnetic field was swept from 0 Oe to 600 Oe. Throughout the experiment, we used a microwave power of 1 mW to ensure consistent power levels for accurate measurements.
\begin{figure}[h!]
\centering
\includegraphics[width=0.7\textwidth]{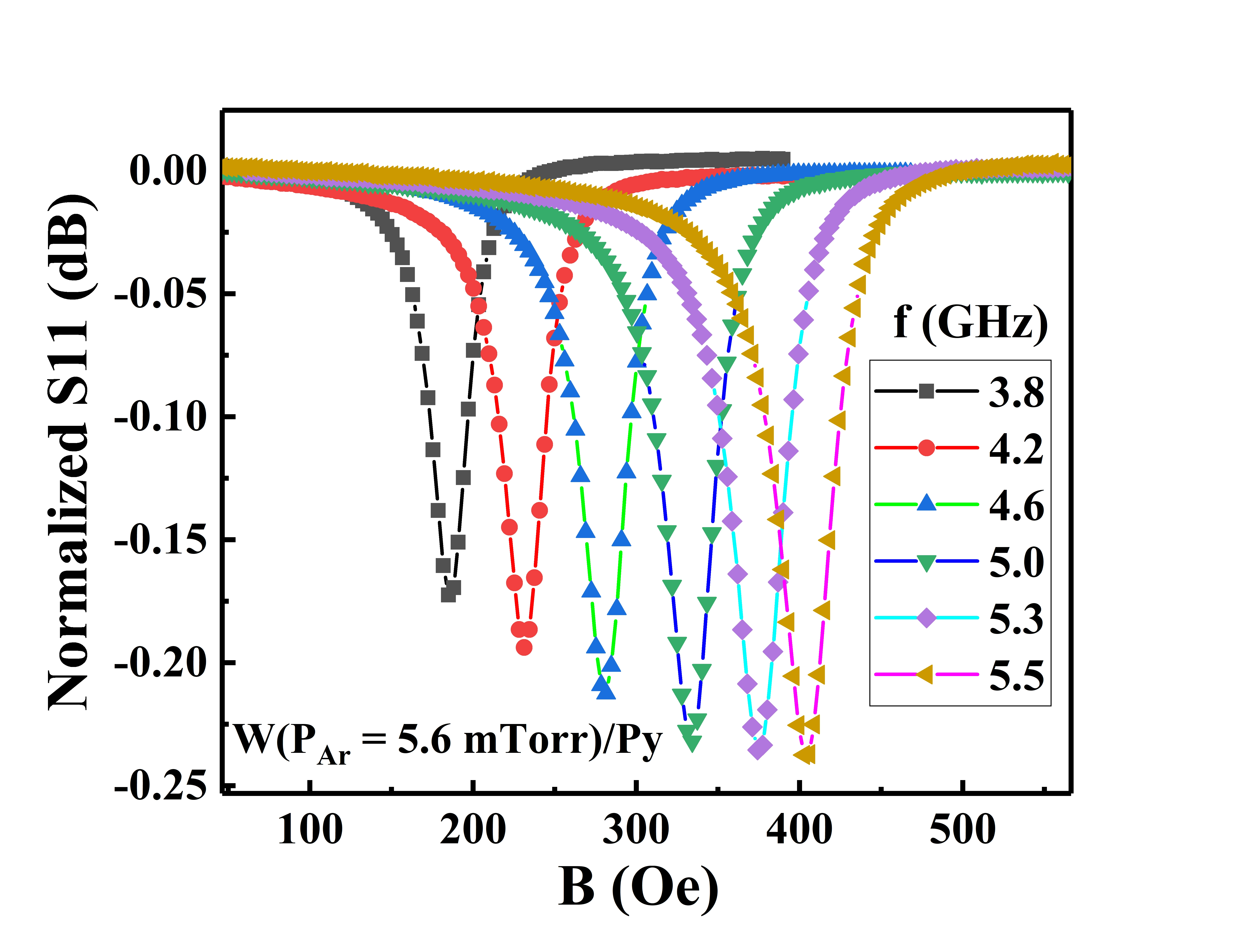}
\caption{The normalized absorption data in $\beta$-W ($P_{Ar}$ = 5.6 mTorr)/Py (set A) as a function of the applied field (B) at f = 3.8, 4.2, 4.6, 5.0, 5.3 and 5.5 GHz at room temperature.}
\label{Figure3} 
\end{figure}

\subsection{Results and Discussion}
\begin{figure*}
\centering
\includegraphics[width=1.0\textwidth]{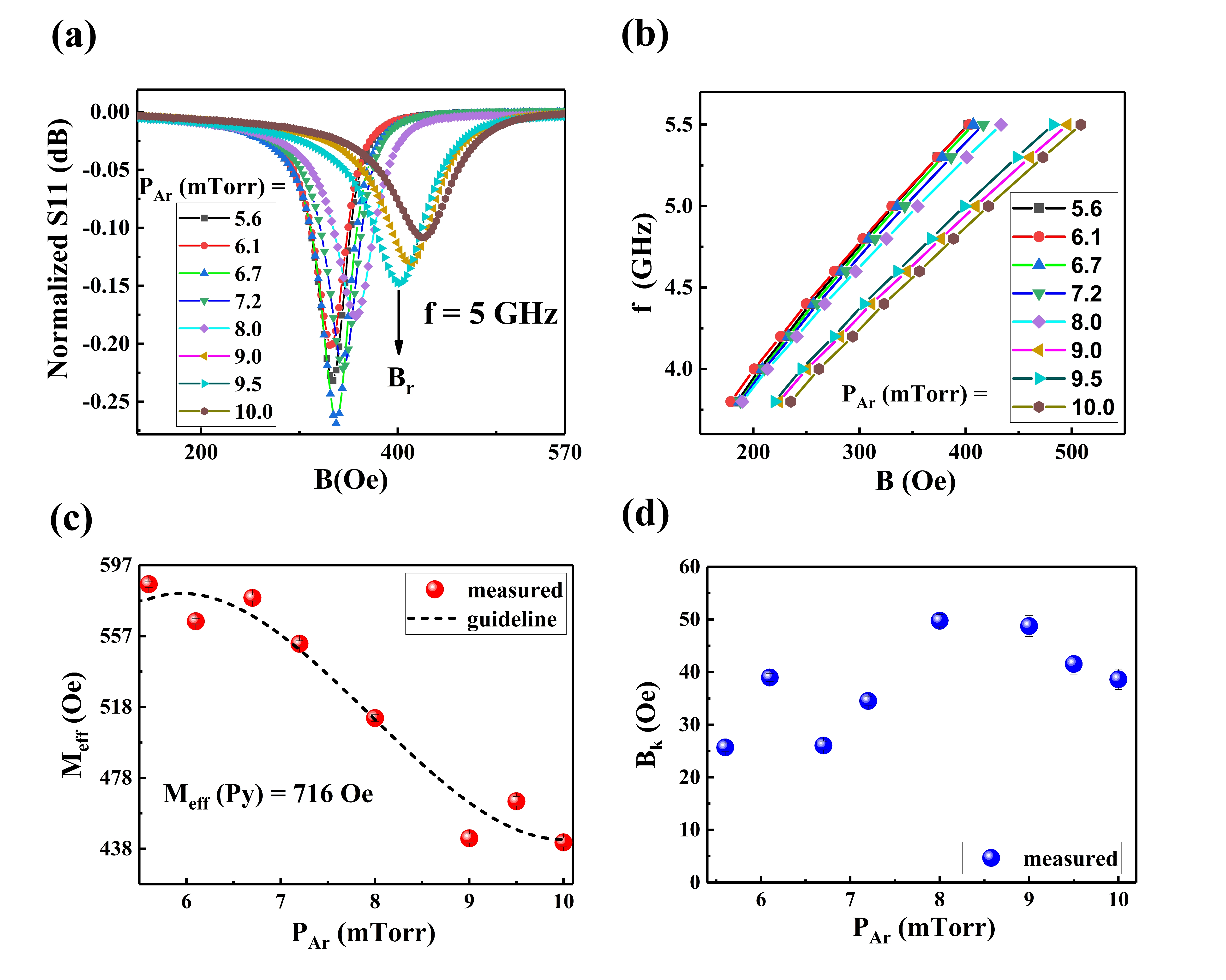}
\caption{(a) Normalized absorption spectra in $\beta$-W($P_{Ar}$)/Py (set A) bilayers as a function of $B$ at a constant f = 5 GHz. $P_{Ar}$ varies from 5.6 mTorr to 10 mTorr, (b) Variation of frequency (f) with applied field (B) which can be fitted with Kittel equation (Eq. \ref{eq:4}). The Solid symbols are used to depict the experimental data points, whereas the solid lines represent the fitted data, (C) $M_{eff}$ vs $P_{Ar}$ (black dotted line serves as a guideline), and (d) The curve $B_{k}$ vs $P_{Ar}$.}
\label{Figure4}   
\end{figure*}
Crystal structure of deposited W films (8 nm) at different $P_{Ar}$ are characterized by x-ray diffraction spectroscopy (XRD) with a resolution of $2^{0}$/min illustrated in  Fig. \ref{Figure2}(a). All deposited W thin films are single-phase $\beta$-W which is A15 type of crystal structure with peaks situated at (200), (210), and (211) corresponding to 2$\theta$ = $35.5^{0}, 40^{0}, 44^{0}$ respectively \cite{hao2015beta,lee2016growth}. The W films grown at monotonically increasing $P_{Ar}$ show a systematic broadening in XRD peaks with lower intensities, which indicates that W films are less crystalline in nature and contain smaller crystallites as described in \ref{Figure2}(a). XRD data provides a rough estimation of the grain size of W films deposited at different $P_{Ar}$. Scherrer's equation \cite{scherrer} can be used to calculate the average grain size of W films, $g = K\lambda/\delta cos(\theta)$, where $K$ is a shape factor ($\approx$ 0.9), $\delta$ is the width of a diffraction peak at half maxima, $\lambda$ is the x-ray wavelength, and $\theta$ is Bragg's angle \cite{maqbool2005surface}. Using the XRD data, we determine grain size as a function of $P_{Ar}$. As $P_{Ar}$ increases from 5.6 mTorr to 10 mTorr, grain size falls from 8.4 nm to 5.3 nm as shown in Fig. \ref{Figure2}(b) inset. The Reduction in grain size as a function of $P_{Ar}$ indicates a higher probability of electron scattering at grain boundaries \cite{Choi2014_es}. Previous studies have reported that $\beta$-W films do have remarkably high resistivity due to the significant electron-phonon interaction presented in the A15 crystal structure \cite{pai2012spin}. According to the reports, the introduction of oxygen inside the vacuum chamber during deposition stabilizes W films in its  $\beta$ phase \cite{karabacak2003beta,salmon2013structure}. The probability of oxygen being incorporated into the film is dependent upon the deposition rate. Films grown at higher $P_{Ar}$ tend to have lower deposition rates, and have a greater tendency to trap more oxygen during deposition. Consequently, films deposited at higher $P_{Ar}$ exhibit increased porosity and disorder resulting in a higher oxygen content which leads to an increase in resistance. As illustrated in Fig. \ref{Figure2}(b), we employ van der Pauw technique for further characterization of resistivity ($\rho_{W}$) at different $P_{Ar}$. A non-monotonic relation between $P_{Ar}$ and $\rho_{W}$ is obtained which can be separated into two segments. Initially, as $P_{Ar}$ increases, $\rho_{W}$ experiences a gradual and slow increase. However, beyond $P_{Ar}$ = 7 mTorr, $\rho_{W}$ exhibits a rapid increase which follows a linear trend until it reaches a saturation point at 10 mTorr with $\rho_{W}$ approaches $\approx$ 1400 $\mu \Omega$-cm. To assess the surface morphology of films, we utilize semicontact topography mode in AFM with a minimum scanning area of 2 $\mu$m $\times$ 2 $\mu$m. The analysis indicates that the films at higher $P_{Ar}$ are more porous, which contributes to higher surface roughness (Fig. \ref{SM_Figure2}). Conversely, when $P_{Ar}$ decreases, the films tend to become smoother implying a compact structure that can be compared with the XRD graph (Fig. \ref{Figure2}(a)). The average surface roughness is observed to decrease from 0.8 nm to 0.3 nm as $P_{Ar}$ decreases from 10 mTorr to 5.6 mTorr.\\ 

Fig. \ref{Figure3} represents the normalized FMR absorption spectra as a function of external field B for $\beta$-W ($P_{Ar}$ = 5.6 mTorr)/Py(15 nm) bilayer over a frequency (f) range from 3.5–5.5 GHz at room temperature. The experimental results are fitted using the Lorentz equation \cite{celinski1997_Lorentzfit}. 

\begin{figure*}
\centering
\includegraphics[width=1\textwidth]{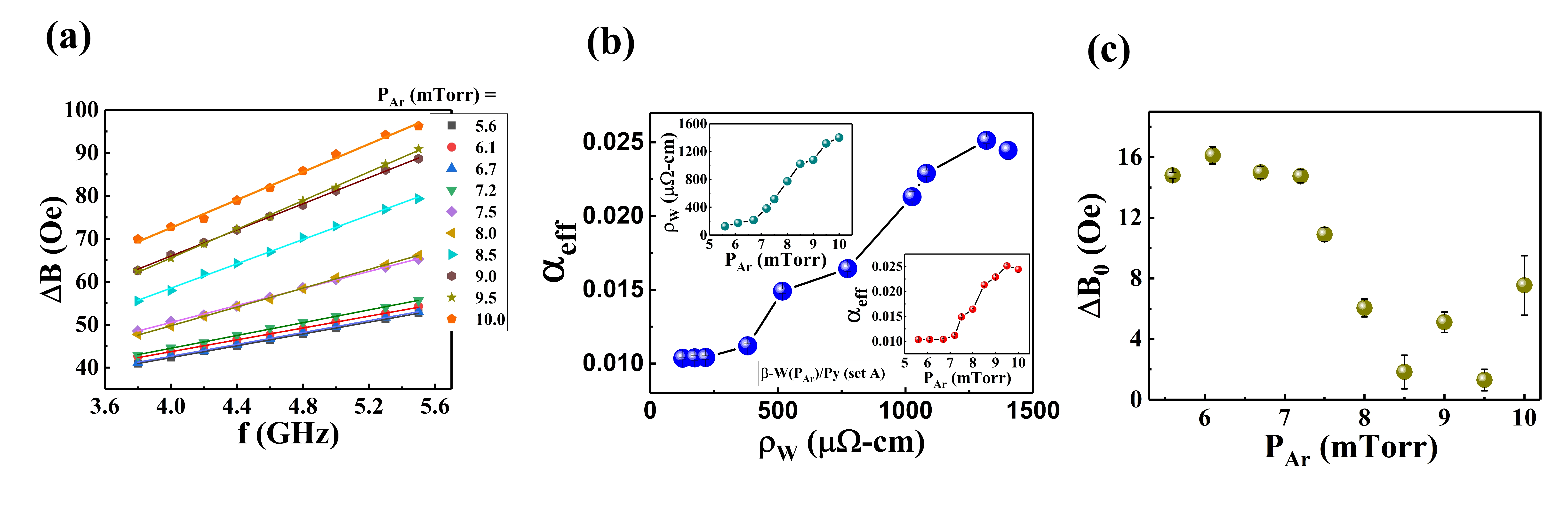}
\caption{(a) Variation of linewidth $\Delta B$ as a function of f in $\beta$-W($P_{Ar}$/Py) (set A) bilayers. The solid symbols are used to indicate the experimental data points, whereas the data points are fitted with Eq. \ref{eq:5} described by the solid lines, (b) $\alpha_{eff}$ calculated from Eq. \ref{eq:5} as a function of $\rho_{W}$ [inset: variation of $\rho_{W}$ with $P_{Ar}$ (dark cyan symbols) and  $\alpha_{eff}$ as a function of $P_{Ar}$ (red solid symbols)], (C) Dependence of $\Delta B_{0}$ on $P_{Ar}$.}
\label{Figure5}   
\end{figure*}
\begin{equation}\label{eq:3}
S_{11} \propto S_{0} \frac{\Delta B}{[4(B-B_{r})^{2}+(\Delta B)^{2}]}
\end{equation}
where $S_{0}$ is the coefficient of absorptive microwave power, B is the in-plane external magnetic field, $\Delta B$  is the absorption linewidth, and $B_{r}$ is the resonance field. The observed increase in $B_{r}$ increases with microwave frequency f which is consistent with the expected behavior \cite{jamali_2013_kittel}. We investigate systematically the FMR spectra of $\beta$-W ($P_{Ar}$)/Py bilayers (set A), with $P_{Ar}$ varying from 5.6 mTorr to 10 mTorr, which corresponds to a range of $\rho_{W}$ from 127 $\mu \Omega$-cm to 1400 $\mu \Omega$-cm. This variation in resistivity permits the study of the dependence of the $\alpha_{eff}$ on $\rho_{W}$ of $\beta$-W thin films. The normalized FMR absorption signal of $\beta$-W ($P_{Ar}$)/Py(15 nm) observed at f = 5 GHz is depicted in Fig. \ref{Figure4}(a). A gradual increase in $B_{r}$ with $P_{Ar}$ is observed, and the absorption curves become broadened, leading to an increase in $\Delta B$. Fig. \ref{Figure4}(b) shows the f vs B curves for $\beta$-W films grown at different $P_{Ar}$. Using Kittel equation \cite{kittel1948theory,jamali_2013_kittel}, the effective magnetization ($ M_{eff}$) and the anisotropy field ($B_{k}$) are extracted from f vs B curve as shown in Fig. \ref{Figure4}(c) and Fig. \ref{Figure4}(d) respectively (set A).

\begin{equation}\label{eq:4}
f= \frac{g\mu_B}{h}\sqrt{(B_{r}+B_{k})(B_{r}+B_{k}+4\pi M_{eff})}   
\end{equation}

With, $g$ (=2.15), $\mu_{B}$, and $h$ are the Lande $g$ factor, Bohr magneton, and Planck's constant, respectively. Fig. \ref{Figure4}(c) illustrates the variation in $M_{eff}$ as a function of $P_{Ar}$, where a monotonous decrease in $M_{eff}$ with increasing $P_{Ar}$ is observed. The evaluated $M_{eff}$ in $\beta$-W ($P_{Ar}$)/Py bilayers is less than the $M_{eff}$ in single-layer Py with the same thickness. The decrease in $M_{eff}$ could be the result of the d-d hybridization at the interface between Py and $\beta$-W \cite{wilhelm_2000_hybridization_PhysRevLett.85.413,pou_2002_hybridization}. Also, the magnetic dead layer created by sp-d hybridization at the interface between Py and $\beta$-W is a potential cause for the decrease in the $M_{eff}$ \cite{shin_1998_dead_layer}. 
Fig. \ref{Figure4}(d) shows the variation in  $B_{k}$ with $P_{Ar}$, which does not have a strong correlation with resistivity of $\beta$-W films and is nearly constant even at higher $P_{Ar}$.\\

\begin{figure}[h!]
\centering
\includegraphics[width=0.5\textwidth]{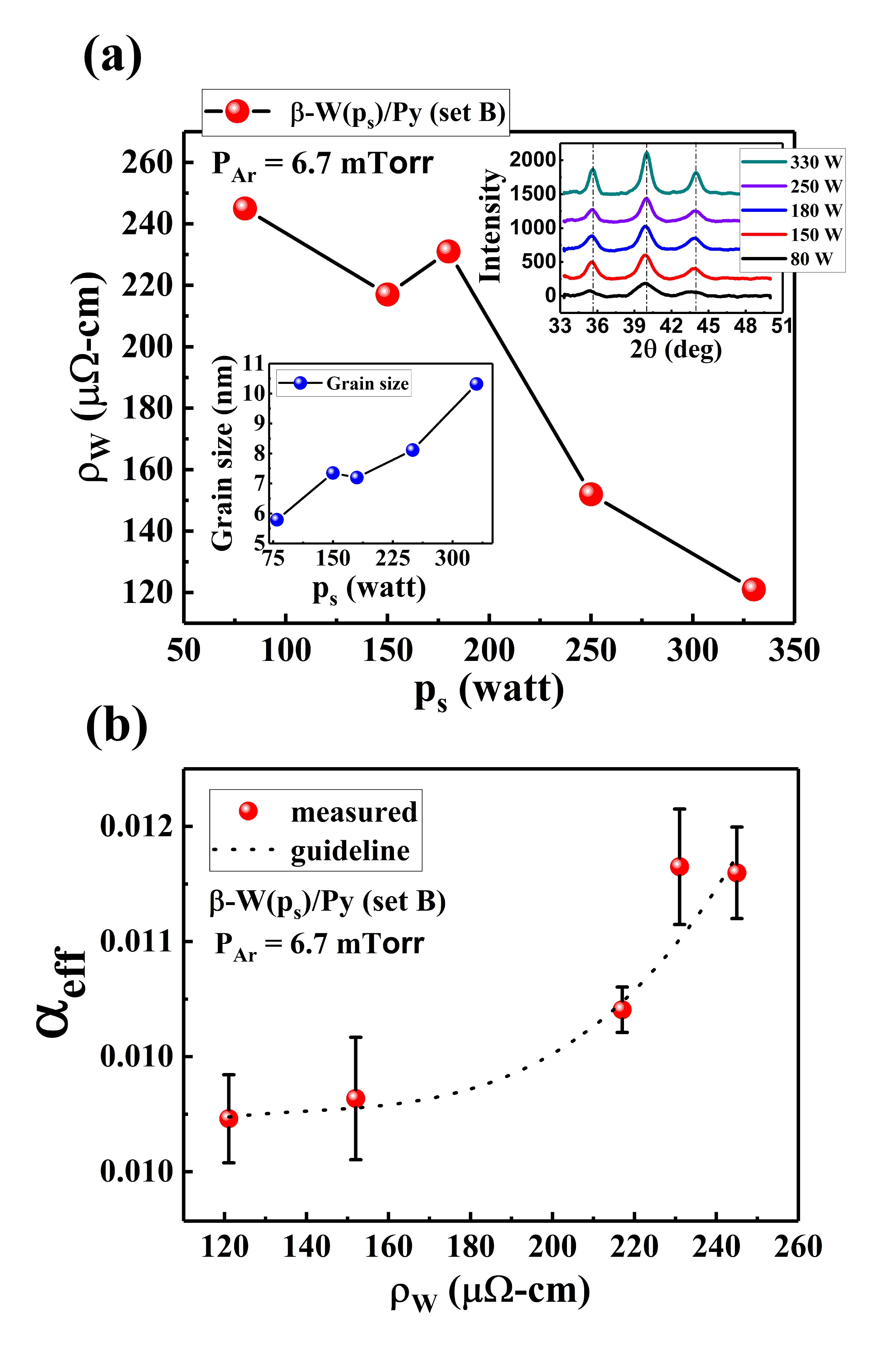}
\caption{(a) Variation in resistivity ($\rho_{W}$) with $p_{S}$ in $\beta$-W (8nm, $p_{S}$)/Py(15 nm) bilayers (set B) (inset: Grain size of $\beta$-W films as a function of $p_{S}$. XRD of $\beta$-W thin films with various $p_{S}$. The peak positions with their relative peak intensities corresponding to the $\beta$ phase of W are shown as vertical dotted lines, (b) $\alpha_{eff}$ is plotted as a function of $\rho_{W}$.}
\label{Figure6}   
\end{figure}

$\Delta B$ as shown in Fig. \ref{Figure5}(a) as a function of f (set A), can be described by a linear relationship \cite{rossing1963}
\begin{equation}\label{eq:5}
\Delta B= \frac{2h\alpha_{eff}}{g\mu_{B}}f + \Delta B_{0}
\end{equation}
where, 1st term is linearly dependent on f and the 2nd term $\Delta B_{0}$ is independent of f known as inhomogeneous line broadening. As described in Fig. \ref{Figure5}(a), the data are represented by solid symbols, and the curves are fitted using Eq. \ref{eq:5}, represented by the solid lines. All $\beta$-W ($P_{Ar}$)/Py samples exhibit a linear response in $\Delta B$ vs f indicating that the damping mechanism is dominated by both the intrinsic Gilbert damping and spin pumping \cite{charilaou_2010_intrinsic,barati2014,zhao2016experimental}. $\Delta B_{0}$ is the frequency-independent extrinsic contribution of damping that reflects the sample's quality during the fabrication process and may be the result of structural imperfections, impurities, or sample roughness \cite{heinrich_1985_linewidth}. Fig. \ref{Figure5}(c) demonstrates that $\Delta B_{0}$ varies from approximately 16 Oe to 1 Oe with $P_{Ar}$. Earlier it has been shown that the roughness of $\beta$-W films increases with $P_{Ar}$ and an opposite trend is observed in $\Delta B_{0}$ which suggests that $\Delta B_{0}$ does not arise from sample roughness. $\alpha_{eff}$ can be evaluated from the plot of $\Delta B$ vs f as shown in Fig. \ref{Figure5}(b). $\alpha_{eff}$ has two components: (a) the intrinsic Gilbert damping ($\alpha_{b}$) caused by the Py layer and (b) damping ($\alpha_{SP}$) due to spin current, pumped into the $\beta$-W films by the process of dissipation \cite{foros2005}. Here, the enhancement in $\alpha_{eff}$ with $\rho_{W}$ is caused by the spin current-induced spin pumping mechanism from Py into $\beta$-W films. Initially, $\alpha_{eff}$ remains constant with a magnitude $\approx$ 0.010 until $\rho_{W}$ reaches $\approx$ 350 $\mu \Omega$-cm, as shown in Fig. \ref{Figure5}(b). Beyond that, $\alpha_{eff}$ begins to increase rapidly, reaching $\approx$ 0.024 at $\rho_{W}$ = 1400 $\mu \Omega$-cm. The increased disorder states at higher $P_{Ar}$ increase $\rho_{W}$, which aids in quicker relaxation and ultimately enhances $\alpha_{eff}$. We conduct a similar study, as depicted in Fig. \ref{Figure6}, in which we vary the input power $p_{s}$ during sputtering while maintaining a constant $P_{Ar}$ (set B). Fig. \ref{Figure6}(a) presents $\rho_{W}$ as a function of $p_{S}$ for $\beta$-W(8nm, $p_{s}$)/Py(15 nm) bilayers with $P_{Ar}$ = 6.7 mTorr. Both XRD data and grain size of $\beta$-W employed in these bilayers are illustrated in the insets in Fig. \ref{Figure6}(a). The relative increase in peak sharpness in XRD indicates that the $\beta$-W films tend to become more crystalline as $p_{s}$ increases. The grain size grows with $p_{s}$ as the films become more smoother, while $P_{Ar}$ remains constant. As shown in Fig. \ref{Figure6}(a), it can be observed that $\rho_{W}$ reduces with $p_{S}$, however, the change is not as significant as what was previously observed when varying $P_{Ar}$ [Fig. \ref{Figure2}(b)]. Fig. \ref{Figure6}(b) depicts the change in $\alpha_{eff}$ with $\rho_{W}$ where $\rho_{W}$ is governed by $p_{S}$. Here, we observe a gradual enhancement in $\alpha_{eff}$ as $\rho_{W}$ increases which can be compared qualitatively to Fig. \ref{Figure5} (b) in the low $\rho_{W}$ range.
Therefore, based on the observation of $\alpha_{eff}$ dependence on both $P_{Ar}$ and $p_{S}$, we may infer that there is a direct relationship between the $\rho_{W}$ of $\beta$-W and the effective damping $\alpha_{eff}$ which defines the spin pumping mechanism in $\beta$-W/Py bilayers.\\
  
\begin{figure}[h!]
\centering
\includegraphics[width=0.5\textwidth]{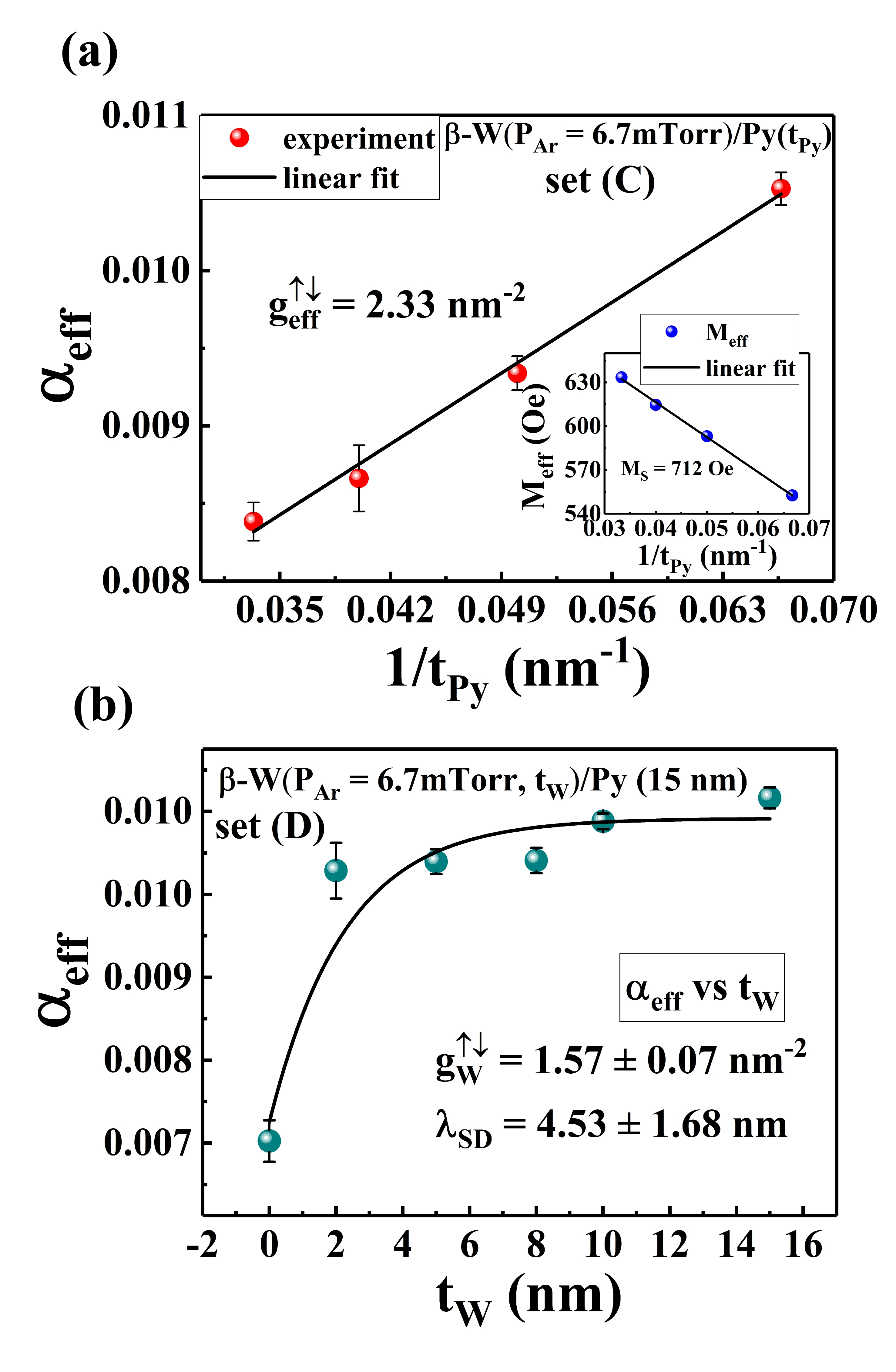}
\caption{(a) Py thickness dependence of $\alpha_{eff}$ (set C), fitted using Eq. \ref{eq:6} with solid red symbols expressing the experimental data and black solid lines are fitted data (inset: $M_{eff}$ vs $1/t_{Py}$), (b) $\alpha_{eff}$ as a function of $t_{W}$ (set D). The curve is fitted using Eq. \ref{eq:9} with solid symbols representing the experimental data, whereas the black solid line describes the fitted data.}
\label{Figure7}   
\end{figure}

\begin{figure*}
\centering
\includegraphics[width=1.0\textwidth]{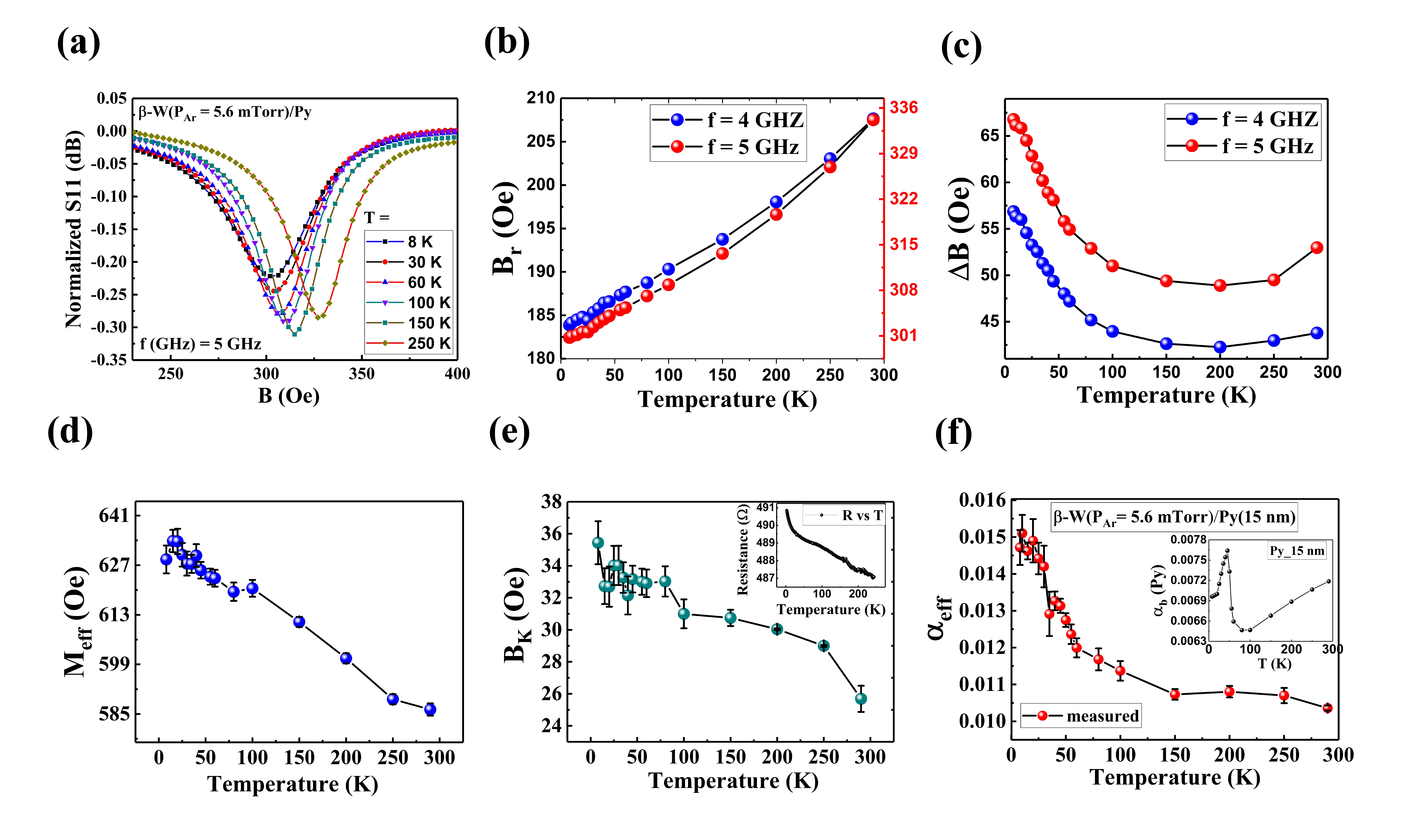}
\caption{(a) Normalized absorption spectra in $\beta$-W($P_{Ar}$ = 5.6 mTorr)/Py (15nm) bilayer as a function of $B$ at T = 8, 30, 60, 100, 150, and 250 K (f = 5 GHz), (b) $B_{r}$ as a function of T at f = 4 and 5 GHz, (c) $\Delta B$ as a function of T at f = 4 and 5 GHz, (d), (e) Temperature dependence of $M_{eff}$ and $B_{k}$ respectively (inset: resistance R vs T of $\beta$-W($P_{Ar}$ = 5.6 mTorr), (f) $\alpha_{eff}$ as a function of temperature (inset: $\alpha_{b}$ vs T of single layer bulk Py).}
\label{Figure8}   
\end{figure*}
In order to determine the interface contribution of $\beta$-W/Py bilayer on spin pumping, we need to evaluate the $g^{\uparrow\downarrow}_{eff}$ by studying the thickness dependence of Py layer as shown in Fig. \ref{Figure7}(a). We use $\beta$-W ($P_{Ar}$ = 6.7 mTorr)/Py ($t_{Py}$) bilayers labeled as set C for our experiment. $\alpha_{eff}$ varies linearly with $1/t_{Py}$ and follows the equation \cite{barati2014,MIZUKAMI2001,Tserkovnyak2002}
\begin{equation}\label{eq:6}
\alpha_{eff} = \alpha_{b} + g\mu_{B}\frac{g^{\uparrow\downarrow}_{eff}}{4\pi M_{S}}\frac{1}{t_{Py}}
\end{equation}
where, $M_{S}$ is saturation magnetization, which can be calculated using a linear equation between $M_{eff}$ and $1/t_{Py}$
\begin{equation}\label{eq:7}
\mu_{0}M_{eff} = \mu_{0}M_{S} - \frac{2K_{S}}{M_{S}d} 
\end{equation}
Eq. \ref{eq:6} is a direct link between the interface and bulk regime, where, the first term $\alpha_{b}$ is intrinsic Gilbert damping of bulk Py and the 2nd term is spin pumping $\alpha_{SP}$ (=$\frac{\delta_{S}}{t_{Py}}$), where, $\delta_{S}$ = $g\mu_{B}\frac{g^{\uparrow\downarrow}_{eff}}{4\pi M_{S}}$ represents the additional damping brought on by combined contribution of Py/vacuum and Py/W interface. The enhanced damping in $\beta$-W/Py of thinner Py films indicates the dominance of surface or interface contribution. The extracted parameters are $\alpha_{b}$ = 0.0061, $\delta_{S}$ = 0.0625 nm in $\beta$-W ($P_{Ar}$ = 6.7 mTorr)/Py ($t_{Py}$) bilayers. We have calculated $M_{S}$ = 712 Oe [inset of Fig. \ref{Figure7}(a)] using Eq. \ref{eq:7}, from which we obtain $g^{\uparrow\downarrow}_{eff}$ = 2.33 $nm^{-2}$ which is comparable with the reported value recently observed experimentally in $\beta$-W \cite{behera2017capping}. The observed value of $g^{\uparrow\downarrow}_{eff}$ is quite high and it confirms a substantial spin transfer via pumping mechanism \cite{deorani_2013_g_mix,tokac_2015_g_mix,azzawi_2016_PhysRevB.93.054402}. We need to determine the $g^{\uparrow\downarrow}_{W}$ in order to comprehend the damping mechanism through the $\beta$-W layer. In a $\beta$-W/Py bilayer, due to interfacial non-equilibrium accumulation, spin back-flow ($J^{back}_{S}$) occurs which further reduces the spin pumping mechanism. Effective spin current due back-flow can be expressed as $J_{S(eff)}$ = $J_{S}(1-e^{-2t_{W}/\lambda_{SD}})$ \cite{Tserkovnyak2002,jiao2013}. The dependence of $\alpha_{eff}$ on $t_{W}$ can be described by the exponential function \cite{Tserkovnyak2002,foros2005,shaw2012}
  
\begin{equation}\label{eq:9}
\alpha_{eff} = \alpha_{b} + g\mu_{B}\frac{g^{\uparrow\downarrow}_{W}}{4\pi M_{S}}\frac{1}{t_{Py}}(1-e^{-\frac{2t_{W}}{\lambda_{SD}}})
\end{equation}
where, $\alpha_{b}$ represents the damping of Py without $\beta$-W layer. Fig. \ref{Figure7}(b) shows the fitted experimental data of $\alpha_{eff}$ on $t_{W}$ using Eq. \ref{eq:9}. It shows that by adding the $\beta$-W layer, the effective damping increases from the bulk Py damping and then saturates after $t_{W}$ $\approx$ 5 nm. The extracted parameters are $g^{\uparrow\downarrow}_{W}$ = (1.57 $\pm$ 0.07) $nm^{-2}$ and $\lambda_{SD}$ = (4.53 $\pm$ 1.68) nm \cite{hao2015beta}\cite{Wang_2014_lambda}. The calculated value of $g^{\uparrow\downarrow}_{W}$ is comparable to the value found in W/Py bilayer recently \cite{behera2017capping}. Due to spin back-flow in $\beta$-W layer, the evaluated $g^{\uparrow\downarrow}_{W}$ is one order of magnitude lower as compared in Pt/Co bilayer reported in \cite{azzawi_2016_PhysRevB.93.054402}. 
The evaluated $\lambda_{SD}$ is consistent with the measured value in $\beta$-W in \cite{hao2015giant}\cite{cho2015large}. Using the coefficient of effective damping, we evaluate the effective spin current $J_{S(eff)}$ through the $\beta$-W layer from FM due to damping, which is provided as follows \cite{deorani_2013_g_mix}
\begin{equation}\label{eq:8}
J_{S(eff)} = \frac{g^{\uparrow\downarrow}_{eff}\gamma^{2}B^{2}_{eff}\hbar [4\pi M_{S}\gamma +\sqrt{(4\pi M_{S}\gamma)^{2}+4\omega^{2}}]}{8\pi \alpha^{2}_{eff}  [(4\pi M_{S}\gamma)^{2}+ 4\omega^{2}]}
\end{equation}
where, $B_{eff}$ is microwave magnetic field and $\omega$ is frequency respectively. Using the known parameters, the calculated $J_{S(eff)}$ is $7.52\times 10^{-10}$ $J/m^{2}$, which is a factor of ten less than the reported value \cite{Nakayama_2012_j_s}. A possible reason is that the  microwave field strength used in our experiment is much weaker than what has been published.\\

For a better understanding of the underlying mechanism, we now focus on the temperature dependence of the spin pumping in the device $\beta$-W($P_{Ar}$=5.6 mTorr/Py (15nm) (set A). Fig. \ref{Figure8}(a) depicts the normalized absorption data at T = 8, 30, 60, 100, 150, and 250 K (f = 5 GHz). The curves tend to get broadened at low T. Fig. \ref{Figure8}(b) illustrates the temperature dependence of resonance field $B_{r}$ at $f$ = 4 and 5 GHz, evaluated from the Kittel equation \cite{kittel1948theory}. As temperature decreases, we observe a shift in $B_{r}$ towards lower magnetic fields which follows a linear trend across the entire temperature range. However, the temperature dependence can be differentiated into two segments by lowering the temperature, with a steeper slope between 290K and 150K than between 100K and 8K. This decrease in $B_{r}$ can be explained by the increased $M_{eff}$ at low temperatures, as depicted in Fig. \ref{Figure8}(d). A similar trend is observed in $B_{K}$ as shown in \ref{Figure8}(e). Temperature dependence of extracted $\Delta B$ at f = 4 and 5 GHz is depicted in Fig. \ref{Figure8}(c) and $\Delta B$ is expected to increase with f. $\Delta B$ increases gradually at lowering T until it reaches 100K, below which it begins to increase swiftly and continues to do so up to 8K. For greater comprehension, we study $\alpha_{eff}$ vs T, where $\alpha_{eff}$ increases slightly from 290K to 100K and begins to increase more rapidly below 100K. This behavior can be explained using the theoretical torque correlation model \cite{kambersky1976ferromagnetic,Gilmore_2007_torque_correlation,Fahnle_2011_torque_correlation,Martin_2020} and our results are consistent with that. According to this model, during magnetic precession, electron-hole pairs are created that relax via lattice scattering with a characteristic relaxation time $\tau$. There are two kinds of transitions involving electron-hole pairs. While the damping brought on by intraband electron-hole pairs is proportional to $\tau$, the damping brought on by interband transitions is inversely proportional to $\tau$. As temperature decreases, the number of phonons begins to decrease, which enhances the relaxation time $\tau$. If only an intraband transition exists, $\alpha_{eff}$ should increase linearly with decreasing temperature. However, non-monotonic increases suggest that there is a modest contribution from interband transition, while intraband transition dominates the overall damping. As illustrated in Fig. \ref{Figure8}(f), $\alpha_{eff}$ rises gradually when the temperature is lowered up to 150 K, below which it starts to increase more rapidly until it reaches a magnitude $\approx$ 0.015 at T=8K.
This result is compared to the damping coefficient $\alpha_{b}$ of single layer Py as shown in the inset of Fig. \ref{Figure8}(f). The graph shows a decrease in $\alpha_{b}$ with temperature and reaches a minimum at 100K, below which $\alpha_{b}$ starts to increase, reaches a maximum at 45K, and decreases again with lowering the temperature. As described in \cite{zhao2016experimental}, this peak may be the result of spin reorientation on the surface of Py at 45 K. In contrast, $\beta$-W($P_{Ar}$=5.6 mTorr/Py (15nm) bilayer shows no such peak. Therefore, comparing the $\beta$-W/Py bilayer and Py, the enhancement in $\alpha_{eff}$ at low-temperature concludes that there is a significant effect of spin pumping from Py into the $\beta$-W layer. 

\subsection{Conclusion}
In conclusion, we systematically study the Gilbert damping in $\beta$-W/Py bilayers by modulating the $P_{Ar}$, which corresponds to different $\rho_{W}$ while maintaining a constant $p_{S}$ and observe that $\alpha_{eff}$ is enhanced with $P_{Ar}$. We also perform a similar experiment to determine $\alpha_{eff}$ as a function of $p_{S}$, while maintaining $P_{Ar}$. Since the resistivity $\rho_{W}$ does not vary considerably with $p_{S}$, a comparably insignificant change in $\alpha_{eff}$ is observed in those bilayers. Both observations suggest a strong correlation between $\alpha_{eff}$ and $\rho_{W}$, which is caused by the increased disorder accompanied by oxygen incorporation into $\beta$-W films during deposition. These disorders modify the relaxation mechanism in the system, which further enhances $\alpha_{eff}$. The thickness dependence of Py indicates that the damping mechanism is dominated by spin pumping in the aforementioned bilayer devices which is supported by an effective spin mixing conductance  $g^{\uparrow\downarrow}_{eff}$ = 2.33 $nm^{-2}$. The role of $\beta$-W on spin pumping in these devices can be explained by considering the back-flow of spin current and the effective spin current is estimated to be $J_{S(eff)}$ = $7.52\times 10^{-10}$ $J/m^{2}$ which can be further used to investigate the inverse spin Hall effect (ISHE) in $\beta$-W/Py heterostructures. The enhancement of $\alpha_{eff}$ with $t_{W}$ estimates the interfacial spin mixing conductance $g^{\uparrow\downarrow}_{W}$ = 1.57 $nm^{-2}$ and $\lambda_{SD}$ = 4.53 $nm$. Furthermore, temperature dependence in $\beta$-W/Py clearly shows a substantial increase in $\alpha_{eff}$ at low temperature that is explained by the torque-correlation model, and the enhancement is attributed to the spin pumping mechanism from the Py layer into the $\beta$-W. Therefore, we come to the conclusion that $\rho_{W}$ of $\beta$-W can be used as a tuning parameter to regulate the $\alpha_{eff}$ which can be helpful in faster switching in spin dynamics in magnetic heterostructures, opening up the possibilities for optimizing their performances. \\ 

\subsection{Acknowledgement}%
We acknowledge the Ministry of Human Resource Development (MHRD), the Science and Engineering Research Board (SERB)
(Grant No. EMR/2016/007950), the Department of Science
and Technology (Grant No. DST/ICPS/Quest/2019/22) for providing the necessary funding. We also sincerely thank the Ministry of Education of the Government of India, the Department of Science and Technology (DST), and IISER Kolkata for funding the scholarship. Ritam Bannerjee, Kanav Sharma, and Sambhu G Nath all contributed to this effort, for which the authors are grateful.

\bibliography{references}

\clearpage  
\onecolumngrid  
\begin{center}
\textbf{\LARGE Supplementary material for `Modulating Spin Current Induced Effective Damping in $\beta-W/Py$ Heterostructures by a Systematic Variation in Resistivity of the Sputtered Deposited $\beta-W$ films'}

\setcounter{figure}{0}
\renewcommand{\thefigure}{S\arabic{figure}}

\begin{figure}[h!]
\centering
\includegraphics[width=1.0\textwidth]{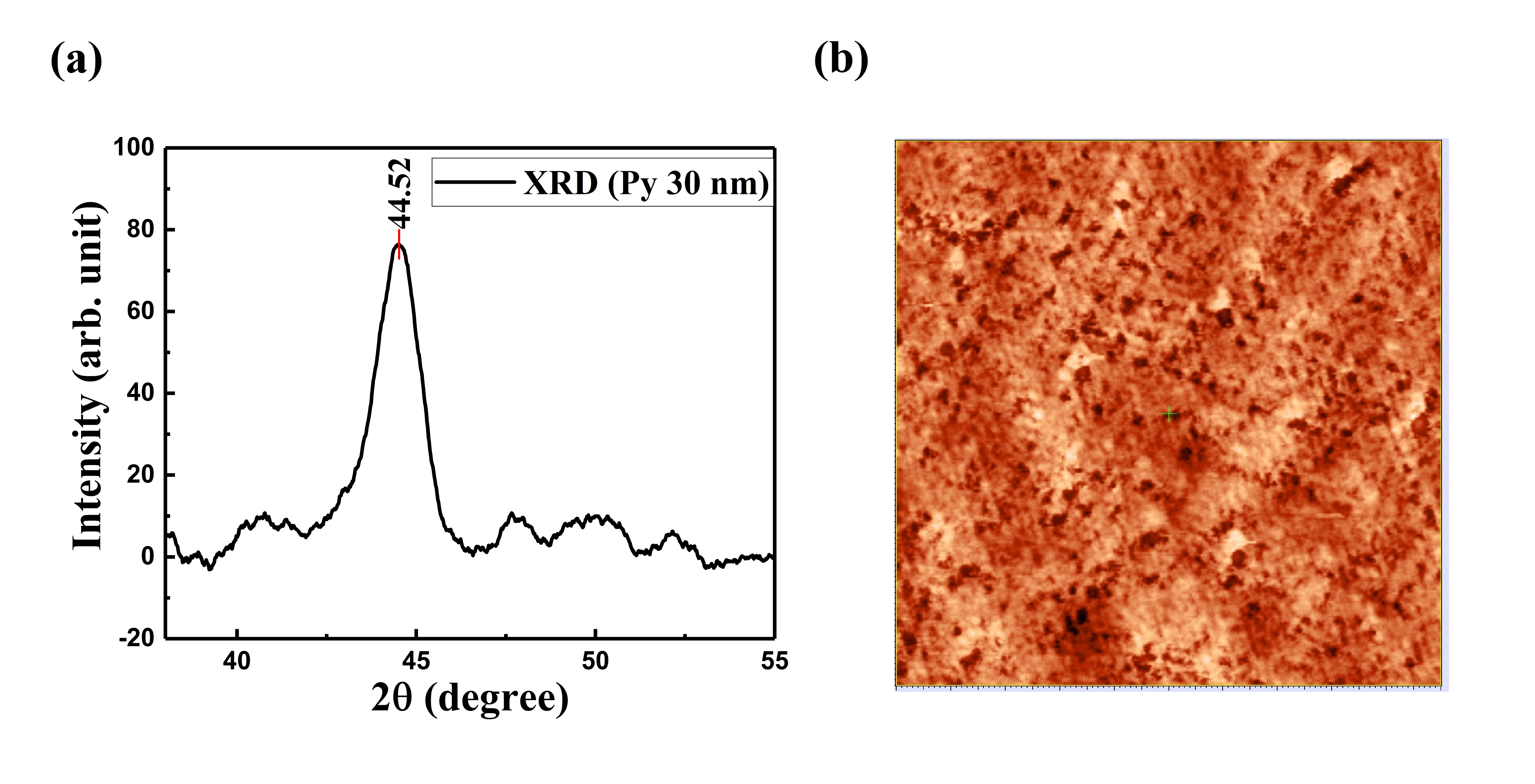}
\caption{(a) XRD of obliquely deposited Py thin film of thickness 15 nm. The peak position at $2\theta = 44.52^{0}$ confirms the fcc crystal structure, (b) AFM image (2 $\mu m$ $\times$ 2 $\mu m$) of 15 nm Py thin film with surface roughness $\approx$ 0.3 nm}
\label{SM_Figure1}   
\end{figure}

\begin{figure}[h!]
\centering
\includegraphics[width=1.0\textwidth]{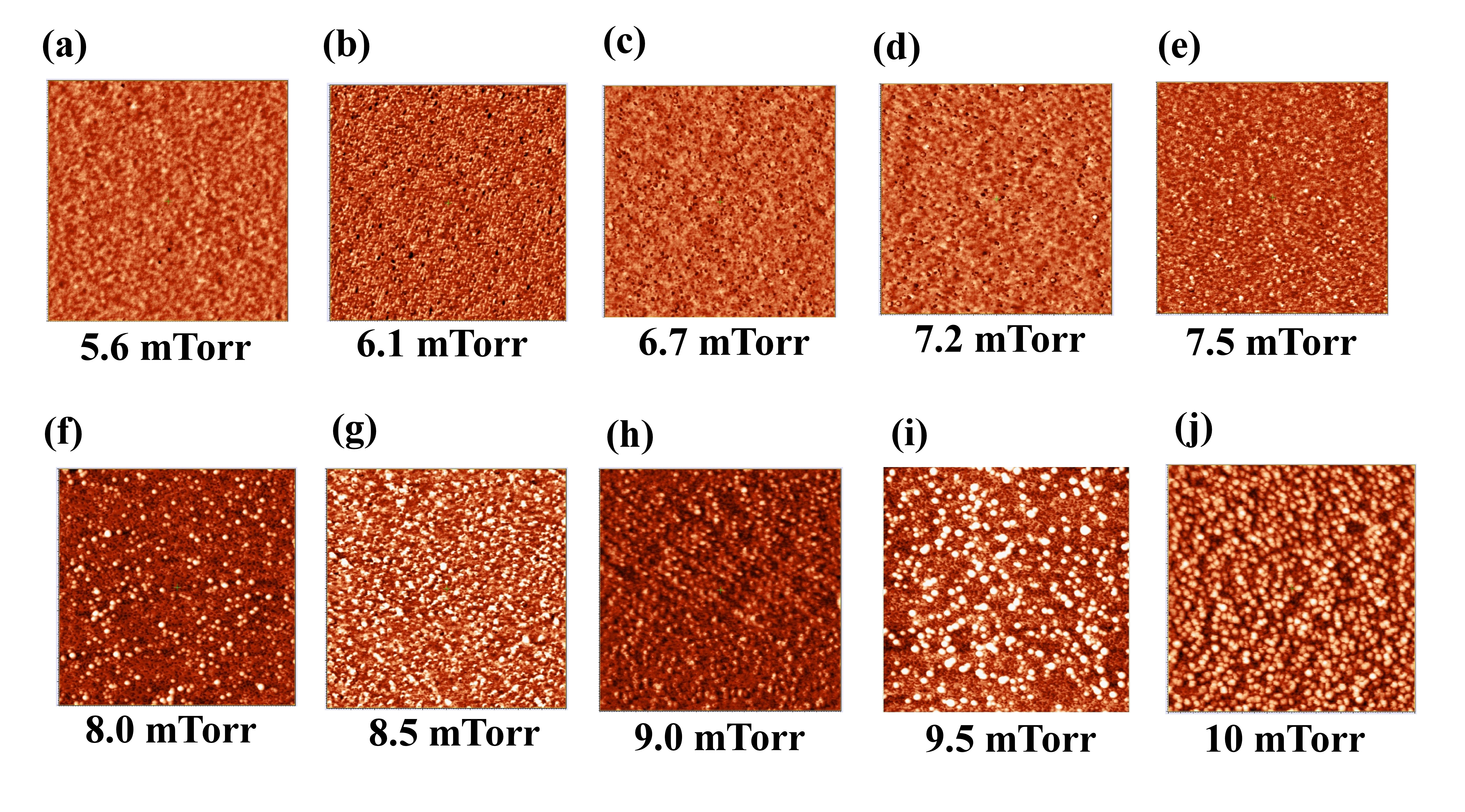}
\caption{(a)-(j) AFM images of $\beta$-W deposited at different $P_{Ar}$ varying from 5.6 mTorr to 10 mTorr, while maintaining a constant $p_{S}$. The surface morphology tends to become rough and less crystalline as $P_{Ar}$ increases and  the grain size decreases accordingly. Average roughness is calculated that increases from 0.3 nm to 0.8 nm (5.6 mTorr to 10 mTorr).}
\label{SM_Figure2}   
\end{figure}

\begin{figure}[h!]
\centering
\includegraphics[width=1.0\textwidth]{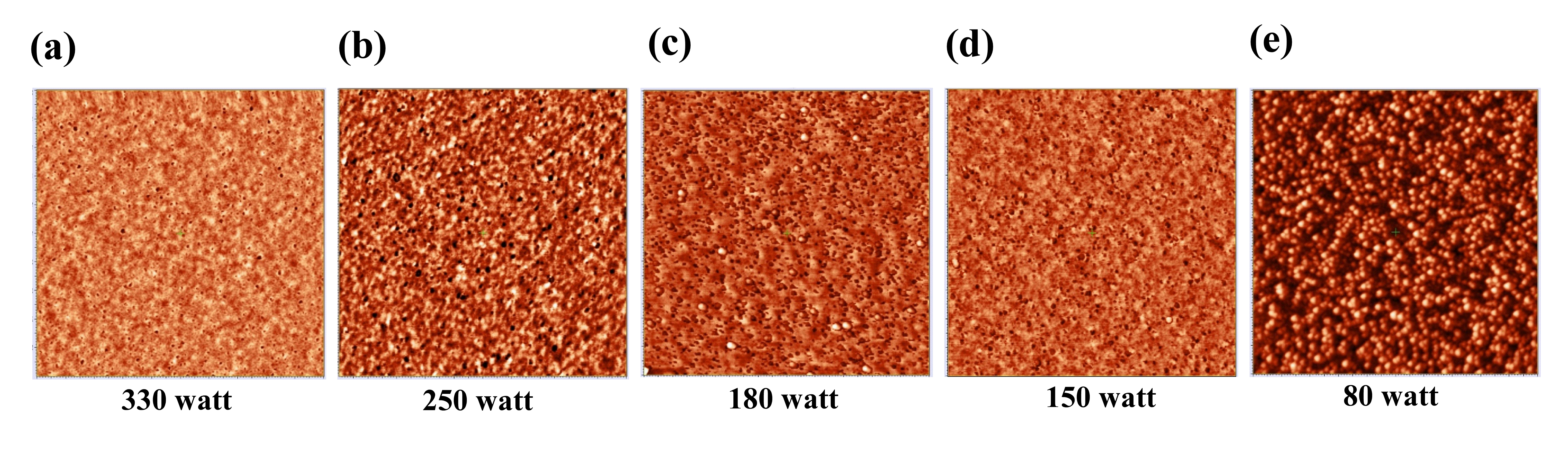}
\caption{a)-(e) AFM images of $\beta$-W deposited at different $p_{S}$ = 330, 250, 180, 150, and 80 watt, while maintaining a constant $P_{Ar}$. Average roughness increases from 0.1 nm to 0.4 nm as $p_{S}$ is lowered from 330 watt to 80 watt.}
\label{SM_Figure3}   
\end{figure}

\end{center}
\end{document}